\documentclass[preprint,onecolumn,nofootinbib]{revtex4}
\usepackage[colorlinks=true,linkcolor=blue,urlcolor=blue,filecolor=black,citecolor=red,pdfstartview=FitV,pdftitle={},pdfsubject={},pdfkeywords={},pdfpagemode=None,bookmarksopen=true]{hyperref}
\usepackage{graphicx}
\usepackage{amsmath}
\usepackage{amssymb,ulem}
\usepackage{color}%
\usepackage{tikz}
\usepackage{dcolumn}
\usepackage{bbold}
\usepackage{times}
\usepackage{pdfsync}
\usepackage{xcolor}
\usepackage{setspace}
\usepackage[T1]{fontenc}

\newcommand{\SOUTHCUT}{
	School of Physics and Optoelectronics, South China University of Technology, Guangzhou 510641,
	People's Republic of China}
\begin{document}
\title{Quasinormal modes of rotating accelerating black holes}
\author{Wei Xiong}
\email{202210187053@mail.scut.edu.cn}
\author{Peng-Cheng Li}
\email{pchli2021@scut.edu.cn, corresponding author}

\affiliation{\SOUTHCUT}
\begin{abstract}

    This paper studies the quasinormal mode spectrum of the scalar perturbation on the background of the rotating accelerating black holes. The quasinormal frequency $\omega$ and the separation constant $\lambda$ are calculated using two methods: the continued fractions method and the directly integration method. The spectrum is found to include three families of modes: the photon sphere modes, the acceleration modes, and the near extreme modes. We investigate the effects of back hole parameters such as spin and acceleration. Empirical formulas are presented for the numerical results, specifically for the acceleration modes in the small black hole limit or the near extreme modes in the extreme black hole limit. An interesting phenomenon known as eigenvalue repulsion is observed when the acceleration modes intersect with the near extreme modes at certain parameter values. The strong cosmic censorship conjecture for spinning C-metric is respected. 

\end{abstract}
\maketitle
    
\section{Introduction}

Black holes (BHs) are compact objects characterized by an event horizon in the universe. Recently, BHs have gained increasing attention in academia, owing to the detection of gravitational waves\cite{LIGOScientific:2016vlm,Brown:2020emp,LIGOScientific:2016aoc}, as well as the first depiction of the BH picture\cite{EventHorizonTelescope:2019dse}. According to the uniqueness theorem \cite{Chrusciel:2012jk}, in four dimensional spacetime, the most general stationary asympotically flat BH solution to the electrovacuum	Einstein field equations is the Kerr-Newman BH \cite{Newman:1965my}. The solution is uniquely characterized by the
mass, angular momentum and the charge. However, the most general solution including more parameters is the Pleba\'nski--Demia\'nski metric in the frame of general relativity\cite{Griffiths:2005qp,Plebanski:1976gy}. This family of metric is of Petrov type D with an aligned electromagnetic field and cosmological constant $\Lambda$ \cite{Griffiths:2005qp,Weir,Kinnersley}. One can reduce the Pleba\'nski--Demia\'nski metric to different solutions with relevant physical interpretations by certain transformations and limiting procedures. 

An intriguing solution called the spinning C-metric belongs to the general Pleba\'nski--Demia\'nski family\cite{Griffiths:2005se,Farhoosh:1980zz,Bicak:1999sa,Hong:2004dm}. It describes a boost-rotation-symmetric stationary spacetime that contains two causally separated rotating BHs accelerating away from each other in opposite spatial directions\cite{Griffiths:2005mi}. This metric is useful for understanding the behavior of moving and accelerating BHs, such as those resulting from a BH superkick\cite{Merritt:2004xa,Bruegmann:2007bri,Gerosa:2016vip,CalderonBustillo:2018zuq}. A superkick may arise from the dynamical processes such as scattering, grazing collision, or merger of two equal-mass BHs \cite{Sperhake:2010uv}. During these events, anisotropic gravitational radiation is emitted, leading to a net emission of linear momentum. As a result, the final remnants undergo gradual acceleration and acquire a recoil velocity.

At the late time stage of such dynamical procedure, the remnant BHs can be described by a perturbed state (e.g., the ringdown stage of the binary BHs merger). Besides, BHs are always perturbed by macroscopic objects or fields in the astronomical environment. The perturbations on the background of BH generate the radiation wave characterized by a set of damped oscillation frequencies, known as quasinormal modes (QNMs) \cite{Kokkotas:1999bd,Berti:2009kk,Konoplya:2011qq,Zhao:2022lrl}. According to the no-hair theorem \cite{Bekenstein:1996pn}, QNMs are only related to the parameters of the background BH, but not to the specific configuration of perturbations. QNMs provide a unique perspective of BH observations and help us to determine the parameters of observed BH \cite{Berti:2005ys, Berti:2018vdi}. 

The QNM spectrum of BH with acceleration and charge, namely the charged C-metric, has been examined in \cite{Destounis:2020pjk,Destounis:2022rpk}. The charge C-metric possesses an acceleration horizon that is analogous to the cosmological horizon of Reissner-Nordstr\"om-de Sitter (RN-dS) BH. This similarity enables the QNM spectrum to include three distinct families of QNMs (the photon sphere modes, the acceleration modes, and the near extreme modes)\cite{Cardoso:2017soq,Cardoso:2018nvb}. The acceleration horizon acts as a boundary that separates the causality between the exterior and interior regions. Imposing the outgoing boundary condition at the acceleration horizon leads to an exponential-law late time behaviors of perturbation for the charged C-metric, instead of a power-law tail under the time domain analysis\cite{Destounis:2020pjk}.

However, realistic BHs in our universe are typically nearly neutral and rotating, with electromagnetic charge quickly neutralized by various mechanisms such as environment plasma, Schwinger pair-creation, or Hawking evaporation \cite{Cardoso:2018nvb}. The spin of BH origins from the collapse of rotating objects \cite{Baumgarte:2016xjw,Gundlach:2016jzm,Gundlach:2017tqq}, the final angular momentum of binary compact object mergers \cite{Berti:2016lat,LIGOScientific:2016lio}, or even accretion of matter onto the BH. 

In this paper, we study the scalar perturbations that are governed by a separable master equation on the background of rotating accelerating BHs. The corresponding QNM spectrum is numerically investigated in full parameter space instead analytically approximated in the Nariai-type near extreme limit\cite{Gwak:2022nsi}. The QNM frequency and the separation constant, which depend on each other, are determined by two methods: the continued fractions method and the directly integration method.

The QNM spectrum includes three distinct families of modes: the photon sphere modes, the acceleration modes, and the near extreme modes. The acceleration modes were previously found in the boosted spacetime\cite{Destounis:2020pjk,Destounis:2022rpk}. When the BH spin becomes large, the photon sphere modes and the near extreme modes branch off from the same set of QNMs\cite{Yang:2012pj,Yang:2013uba}. The photon sphere modes are associated with the peaks of the potential barrier while the near extreme modes are related to the near horizon geometry of the Kerr BH. 

We find an intriguing phenomenon called eigenvalue repulsion, which emerges from the intersection between the acceleration modes and the near extreme modes. The eigenvalue repulsion is a well-established phenomenon in the energy level problems of simple quantum mechanical models with self-adjoint Hamiltonians. This phenomenon within the QNM spectrum governed by non-self-adjoint operators, was first reported by Dias in the cases of Kerr-Newman BH recently \cite{Dias:2021yju,Dias:2022oqm} and (higher dimensional) Kerr-dS BHs \cite{Davey:2022vyx}. For the rotating accelerating BHs, the eigenvalue repulsion arises as a result of the intersection between acceleration modes and near-extreme modes in the complex plane.

The QNMs can also be used to examine the strong cosmic censorship (SCC) conjecture, which is associated with an infinite blueshift of perturbations at the Cauchy horizon\cite{Dafermos:2003wr,Luk:2017jxq}. Recent works demonstrate that the SCC is violated by nearly extreme RN-dS BHs when considering neutral scalar perturbations\cite{Cardoso:2017soq}. The exponential decay of these perturbations can suppress the blueshift amplifications at the Cauchy horizon. However, the violation of SCC is saved by the presence of charged massive scalars\cite{Cardoso:2018nvb} and SCC is respected by perturbations on the background of Kerr-dS BHs \cite{Dias:2018ynt}. In accelerating spacetimes, there also exists a competitive mechanism between the blueshift instability and the exponential decay of scalar perturbations at the Cauchy horizon\cite{Destounis:2020pjk}. The SCC conjecture is violated when considering scalar perturbations on the charged C-metric\cite{Destounis:2020yav} or conformally scalar accelerating BHs with neither charge nor rotation\cite{Zhang:2023yco}. In this paper, we investigate the spinning C-metric for perturbations.

The remainder of this paper is organized as follows. We review the spinning C-metric and the perturbation equations separated from the master equation in section \ref{sectionII}. In section \ref{sectionIII}, the two methods we used are introduced. The results are presented in section \ref{sectionIV}. We give a summary and some discussions in section \ref{sectionV}. By convention, we employ the geometric units $G=c=1$. 

\section{Accelerating black hole}

\label{sectionII}

\subsection{Spinning C-matric}

The spinning C-metric can be obtained by imposing certain constraints on the Pleba\'nski--Demia\'nski metric\cite{Griffiths:2005qp}. One can express the line element corresponding to the spinning C-metric by using the Boyer-Lindquist-type coordinates ($t,r,\theta,\phi$)\cite{Bini:2008mzd}
\begin{eqnarray}
    ds^{2} &=& \frac{1}{\Omega} \Big( \frac{1}{\Sigma}(Q-a^{2}P\sin^{2}\theta) dt^{2} - \frac{2a\sin^{2}\theta}{\Sigma} \big( Q-P(r^{2}+a^{2}) \big) dt d\phi \nonumber \\
           & & -\frac{\sin^{2}\theta}{\Sigma} \big( P(r^{2}+a^{2})^{2} -a^{2}Q\sin^{2}\theta \big) d\phi^{2} - \frac{\Sigma}{Q} dr^{2} - \frac{\Sigma}{P} d\theta^{2} \Big), 
    \label{eq:metric}
\end{eqnarray}
with the definitions of functions
\begin{eqnarray}
    \Omega &=& 1-\alpha r \cos\theta , \ \ \ \Sigma = r^{2}+a^{2}\cos^{2}\theta \nonumber \\
    P      &=& 1- 2\alpha M \cos \theta +a^{2} \alpha^{2} \cos^{2} \theta, \nonumber \\
    Q      &=& \Delta (1-\alpha ^{2} r^{2}), \ \ \ \Delta = r^{2} - 2Mr +a^{2}, 
\end{eqnarray}
where the parameters $M$, $\alpha$, and $a$ are related to the BH mass, acceleration, and rotation respectively. 
The spinning C-metric reduces to the nonrotating vacuum C-metric when $a=0$, to the vacuum Kerr metric when $\alpha=0$, and to the Schwarzschild metric when both $a$ and $\alpha$ vanish.

There are conical singularities at the axis $\theta= 0$ and $\theta = \pi$, indicating the existence of deficit angles. Without loss of generality, we specify $\phi \in [0,2\pi/P(\pi))$ to remove the conical singularity at $\theta = \pi$. However, at the opposite pole ($\theta=0$), the deficit angle persists and it is not possible to remove both conical singularities simultaneously \cite{Destounis:2020pjk}.

According to the expression of $Q$, we have $r_{\pm} = M \pm \sqrt{M^{2}-a^{2}}$ as the outer event horizon and the inner Cauchy horizon and $r_{\alpha} = \alpha ^{-1}$ as the acceleration horizon related three null hypersurfaces of the rotating accelerating BH. Our calculation is limited to the spherical corona $r_{+}<r<r_{\alpha}$ (or $\alpha <\alpha_{ext} \equiv 1/(1+\sqrt{M^{2}-a^{2}})$) by convention. Such constraint gives $Q>0, \, P(\theta)>0$ for $r>0, \,\theta \in [0,\pi]$. 

The surface gravity at each horizon $r_{i} \ \in \ \{ r_{-},r_{+},r_{\alpha} \}$ is given by
\begin{equation}
    \kappa_{i} \equiv \left. \frac{|\partial_{r}Q(r)|}{2(r^{2}+a^{2})} \right|_{r = r_{i}}.
    \label{eq:surface gravity}
\end{equation}

\subsection{Perturbation equation}

A separable master equation, describing the massless field perturbations to the spinning C-metric with any spin $s$, has been established by Bini\cite{Bini:2008mzd} in terms of gauge- and tetrad-invariant quantities. This approach based on the Newman-Penrose formalism was originally developed to study perturbations on the background of the Kerr BH for any spin. Imposing the separable solutions
\begin{equation}
    \varphi(t,r,\theta,\phi) = \Omega^{2s+1} e^{-i\omega t} e^{i m \phi} R(r) \frac{Y(\theta)}{\sqrt{P}}, 
\end{equation}
the master equation can be separated into the radial part
\begin{eqnarray}
    0 &=& Q^{-s} \frac{d}{dr} \left(Q^{s+1} \frac{dR(r)}{dr}\right) + V_{rad}(r) R(r),  \label{eq:radial} \\
    V_{rad} (r) &=& -2r \alpha^{2} (r-M)(1+s)(1+2s) + \frac{((a^{2}+r^{2})\omega-am)^{2}}{Q} \nonumber \\
    && - 2is\left(-\frac{am\partial_{r} Q}{2Q}+\frac{\omega M (r^{2}-a^{2})}{\Delta}-\frac{\omega r \sigma}{1-\alpha^{2}r^{2}}\right) +2\lambda,
\end{eqnarray}
and the following angular part
\begin{eqnarray}
    0 &=& \frac{1}{\sin \theta} \frac{d}{d\theta} \left(\sin\theta \  \frac{d Y(\theta)}{d\theta} \right) + V_{ang}(\theta) Y(\theta) \label{eq:angular} \\
    V_{ang}(\theta) &=& -\frac{2\lambda-s(1-\alpha^{2}a^{2})}{P} + \frac{1}{P^{2}} 
    \Big(
    \frac{ -\sigma^{2} s^{2} \cos^{2}\theta + 2\sigma s w \cos\theta-w^{2}\cos^{2}\theta}{\sin^{2}\theta} \nonumber \\
    && +z^{2} \cos^{2}\theta - 2s\sigma z \cos\theta- (z+w-4s\alpha M)^{2} \nonumber \\
    &&  +\alpha^{2} \left((M^{2}-a^{2})\sin^{2}\theta +4sa^{2} \cos\theta (2s \alpha M-w)\right) 
    \Big),
\end{eqnarray}
where $z \equiv a \omega + s \alpha M$, $w \equiv 2 s \alpha M-m$, and $\sigma \equiv 1+a^{2}\alpha^{2}$. The spin-weight parameter $s$ of the above equations is set to be 0 for the scalar perturbations in this study. The parameter $\omega$ signals the oscillation frequency. To remove the conical singularity along $\theta = \pi$, the azimuthal separation constant $m$ needs to be of the form $m\equiv m_{0} P(\pi)$, where $m_0$ is a positive integer. 

The separation constant $\lambda$, as an eigenvalue of the spin-weighted spheroidal harmonics in the generalized Teukolsky formalism, depends on $\omega$ in the presence of a nonzero BH spin parameter $a$. $\lambda$ can be written in familiar form for certain limit cases of the parameters, e.g., $\lambda = -l(l+1)/2$ with $a=0,\alpha=0$ for Schwarzschild case\cite{Bini:2008mzd}, $\lambda = (-A_{lm}-a^{2}\omega^{2}+2am\omega)/2$ with $\alpha=0$ for Kerr case\cite{Leaver:1985ax}, $\lambda = (1/3-\lambda')/2$ with $a=0$ for nonspinning C-metric case\cite{Destounis:2020pjk,Destounis:2022rpk}. The symbols ($l,A_{lm},\lambda'$) represent the separation constant for corresponding references.

The calculation of QNM frequency $\omega$ is the eigenvalue problems of (\ref{eq:radial}) and (\ref{eq:angular}) related to the QNM boundary condition. Physically, only purely ingoing waves can exist at the event horizon, while at the acceleration horizon the purely outgoing wave uniquely present 
\begin{equation}
    R(r) \sim \left\{
    \begin{array}{ll}
    (r-r_{+})^{A}, \ \  &  r\rightarrow r_{+} \\
    (r_{\alpha}-r)^{B}, \ \  & r\rightarrow r_{\alpha},
    \end{array}
    \right.
    \label{eq:boundary condition radial}
\end{equation}
with $A = -i[\omega-am/(r_{+}^{2}+a^{2})]/2\kappa_{+}$ and $B = -i[\omega-am/(r_{\alpha}^{2}+a^{2})]/2\kappa_{\alpha}$.
We also require the solution to be regular at the interval boundaries of $\theta$
\begin{equation}
    Y(\theta) \sim \left\{
    \begin{array}{ll}
    \left( \frac{\cos\theta-1}{2}\right) ^{C}, \ \  &  \theta \rightarrow 0 \\
    \left( \frac{\cos\theta+1}{2}\right)^{D}, \ \  & \theta \rightarrow \pi
    \end{array}
    \right.
    \label{eq:boundary condition angular}
\end{equation}
where $C=\frac{m}{2-4M\alpha+2a^{2}\alpha^{2}}$ and $D = \frac{m}{2+4M\alpha+2a^{2}\alpha^{2}}$, to avoid the disastrous divergence of solutions at $\theta=0,\pi$. The more evident boundary behavior discussion can be found in \cite{Bini:2008mzd} (also in \cite{Destounis:2020pjk} for non-spinning C-metric) where the perturbation equation was transformed into the Schr\"odinger-like form. 

Remarkably, the separation content $\lambda$ and the QNM $\omega$ are eigenvalues entangled with each other for the two equations with proper boundary conditions (\ref{eq:boundary condition radial}) and (\ref{eq:boundary condition angular}). $\lambda$ also depends on the acceleration $\alpha$ of the accelerating BHs. Our treatment follows the way by Leaver\cite{Leaver:1985ax}, i.e., given a tentative value of $\omega$, we solve $\lambda$ from (\ref{eq:angular}) and then judge whether the corresponding solution of $\omega$ satisfies the QNM conditions (\ref{eq:boundary condition radial}). The judgment of the two methods is similar: the continued fractions equation (\ref{eq:continuedfraction}) tends to be 0 for the continued fractions method and the resulting determinant (\ref{eq:determinant}) tends to be 0 for the directly integration method. This program can be regarded as the root-finding problem of a numerical function $F(\omega)$. A QNM can be found iteratively with a beginning tentative $\omega$ sufficiently close to it. The two methods will be introduced in detail in section \ref{sectionIII}. 

\section{method}
\label{sectionIII}

In our study, we employ the continued fractions technique to evaluate $\lambda$ for angular parts and use both the two methods to determine $\omega$. 
The comparison of the two methods is presented in Sec.\ref{secIV.3}.

\subsection{Continued fractions method}

The well known Teukolsky equation with any spin was solved by the continued fractions method on the background of Kerr BH. In the seminal work by Leaver\cite{Leaver:1985ax}, both the QNM frequency and the separation constant are calculated using the continued fractions method. 

Generally, the continued fractions method applies to most second order linear homogeneous differential equations with variable $z$ defined within the unit circle $0<|z|<1$\cite{Konoplya:2011qq}. The singularities of the equation should be removed outside the unit circle by redefining the coordinates for the convergence of the Frobenius series \footnote{
One can apply the continued fractions method beyond the convergence condition, as discussed in appendix B of \cite{Destounis:2020pjk}.}. We develop a new code, which is a slightly different from Leaver's work. 

Let us start with the construction of the Frobenius series
\begin{eqnarray}
    R(r) &=& \left(\frac{r-r_{+}}{r-r_{-}}\right)^{A} \left( \frac{r_{\alpha}-r}{r-r_{-}} \right)^{B} \sum_{i=0}^{N} a_{i}  \left( \frac{r-r_{+}}{r-r_{-}} \frac{r_{\alpha}-r_{-}}{r_{\alpha}-r_{+}} \right)^{i} \label{eq:fronbenius radial}
    \\
    Y(\theta) &=& \left( \frac{\cos\theta-1}{2}\right) ^{C}\left( \frac{\cos\theta+1}{2}\right)^{D} \sum_{i=0}^{K} b_{i} \left( \frac{\cos\theta+1}{2} \right)^{i}, \label{eq:fronbenius angular}
\end{eqnarray}
where the two series are truncated to order $N$, $K$ respectively. The Frobenius series are imposed the boundary behaviors from the boundary conditions (\ref{eq:boundary condition radial}) for $R(r)$ and (\ref{eq:boundary condition angular}) for $Y(\theta)$. All singular points appearing in equations are removed outside the unit circle through the definition of series. We will introduce the radial part as a template because both two parts share the same code in the following discussion. 

Substituting (\ref{eq:fronbenius radial}) into radial equation (\ref{eq:radial}), in general, one can obtain a $N$-term recurrence relation 
\begin{equation}
    \sum^{\textrm{min}(N-1,i)}_{j=0} \bar{\beta}_{j,i}^{(N)} a_{i-j} =0, \qquad \textrm{for} \; i>0.
    \label{eq:N term recurrence relation}
\end{equation}
The algebraic relation between coefficients $a_{i}$ allows one to reduce (\ref{eq:N term recurrence relation}) to a 3-term recurrence relation
\begin{eqnarray}
    \beta^{(3)}_{0,i}a_{i}+\beta_{1,i}^{(3)}a_{i-1}+\beta_{2,i}^{(3)}a_{i-2}&=&0, \quad \textrm{for} \; i>1, \nonumber \\
    \beta^{(3)}_{0,1}a_{1}+\beta_{1,1}^{(3)}a_{0} &=& 0
    \label{eq:3 term recurrence relation}
\end{eqnarray}
step by step through an iterative program, which $\beta_{i}$ is constituted by $\bar{\beta}_{j}$ as a function of system parameters. The resulting continued fractions equation is given by
\begin{equation}
    \beta^{(3)}_{1,1}-\frac{\beta^{(3)}_{0,1} \beta^{(3)}_{2,2}}{\beta^{(3)}_{1,2}- \frac{\beta^{(3)}_{0,2}\beta^{(3)}_{2,3}}{\beta^{(3)}_{1,3}- \cdots }}=0, 
    \label{eq:continuedfraction}
\end{equation}
while we express $a_{1}/a_{0}$ from the two equations in (\ref{eq:3 term recurrence relation})\cite{Leaver:1985ax}. The equation above holds while $\omega$ is the QNM frequencies.

\subsection{Directly integration method}

The directly integration method has been introduced by Pani\cite{Pani:2013pma} and employed to coupled perturbation equations\cite{Blazquez-Salcedo:2016enn,Pierini:2021jxd,Pierini:2022eim}
or even the perturbations on the numerical solution of hairy BHs \cite{Blazquez-Salcedo:2017txk,Blazquez-Salcedo:2018jnn}. It is a powerful method competent to the decoupled perturbation equations of analytic system in this paper. The main idea is to match two nontrivial solutions from opposite boundaries with corresponding boundary conditions (\ref{eq:boundary condition radial}) at an arbitrary midpoint $r_{m}$. The ingoing solution $R_{\textrm{in}}$ integrating from the event horizon must be proportional to the outgoing solution $R_{\textrm{out}}$ integrating from the acceleration horizon while eigenvalue $\omega$ of solutions is QNM frequency. 

This method begins with constructing the series approximation at boundaries
\begin{eqnarray}
    S_{\textrm{in}}(r)  &=& (r-r_{+})^{A} \sum_{i=0}^{N} c_{i} \  (r-r_{+})^{i}, \ \ \ \textrm{at} \ \ r_{+}, \label{eq:series ingoing} \\
    S_{\textrm{out}}(r) &=& (r_{\alpha}-r)^{B} \sum_{i=0}^{N} d_{i} \ (r_{\alpha}-r)^{i}, \ \ \ \ \textrm{at} \ \  r_{\alpha}.
\end{eqnarray}
These series are truncated to order $N$. Plugging (\ref{eq:series ingoing}) into the radial equation (\ref{eq:radial}) one can solve the expression  for each $c_{i}$ in terms of $c_{0}$. $c_{0}$ is set to a nonvanishing constant (such as $1$ in this paper) for approximating nontrivial solutions. This solved series can be used as an approximate boundary condition of ingoing solution $R_{\textrm{in}}$ near the event horizon
\begin{equation}
    R_{\textrm{in}} (r_{+}+\epsilon) = S_{\textrm{in}} (r_{+}+\epsilon)
\end{equation}
with some small value $\epsilon$. Then $R_{\textrm{in}}$ is integrated from $r_{+}+\epsilon$ to $r_{m}$. The procedure for solving $R_{\textrm{out}}$ is similar. As a result, one can construct the determinant
\begin{equation}
    \textrm{Det} = \left|
    \begin{array}{ll}
        R_{\textrm{in}}(r_{m}) & \partial_{r} R_{\textrm{in}}(r_{m})  \\
        R_{\textrm{out}}(r_{m}) & \partial_{r} R_{\textrm{out}}(r_{m})
    \end{array}    
    \right|.
    \label{eq:determinant}
\end{equation}
The QNMs are obtained by imposing $\textrm{Det}=0$.

\section{Results}

\label{sectionIV}

We investigate the QNM spectrum of rotating accelerating BHs. The mass $M$ is fixed to 1 and hence all the physical quantities below are written as the simple dimensionless forms with $M$ such as $M \omega = \omega$. We only present the mode with the separation constant $\lambda$ with the largest real part (i.e., $l=m_{0}$ for the Schwarzschild case) in this paper.

\subsection{Spectrum}

\begin{figure}[htbp]
    \centering
    \includegraphics[width = 0.4\textwidth]{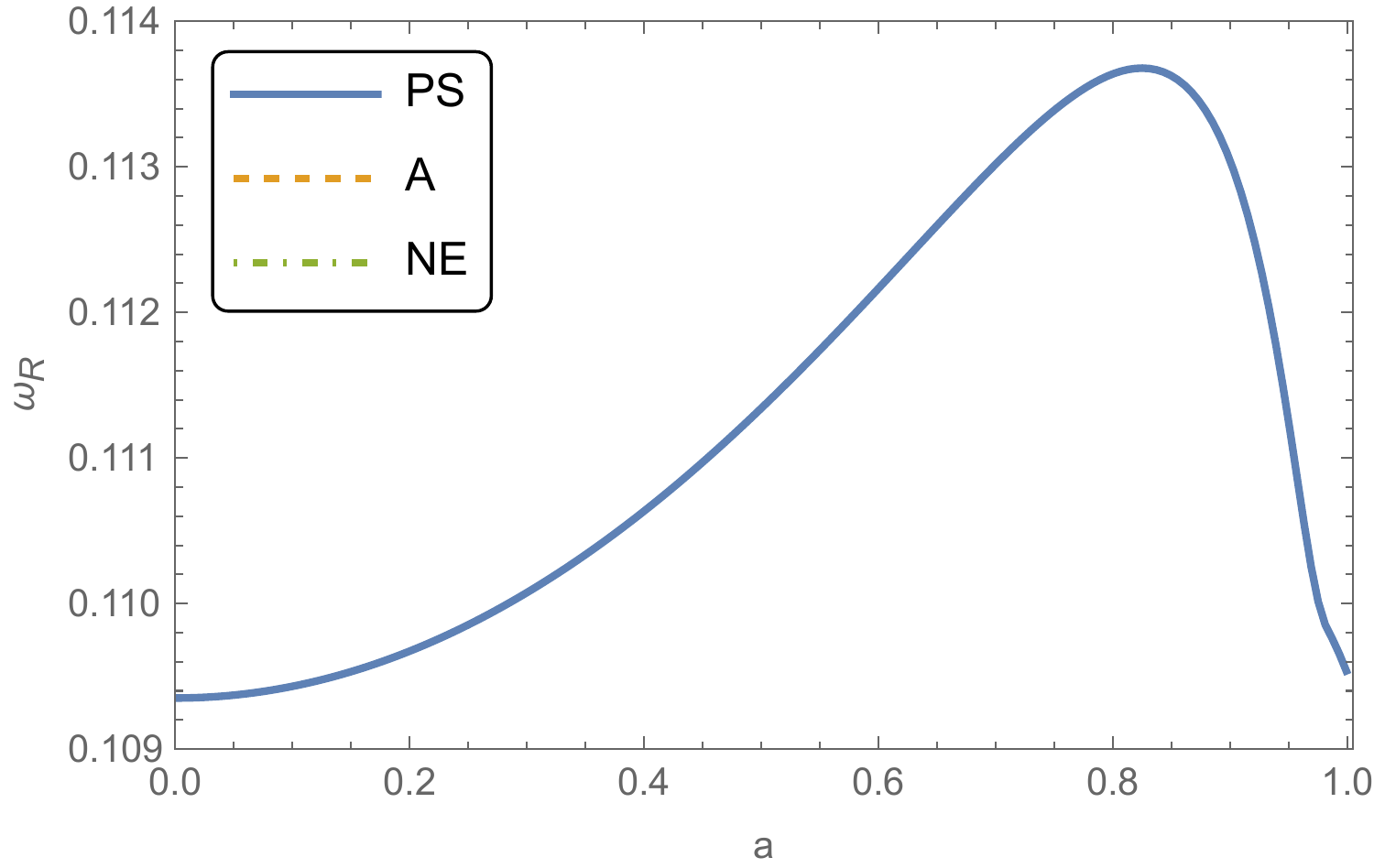}
    \includegraphics[width = 0.4\textwidth]{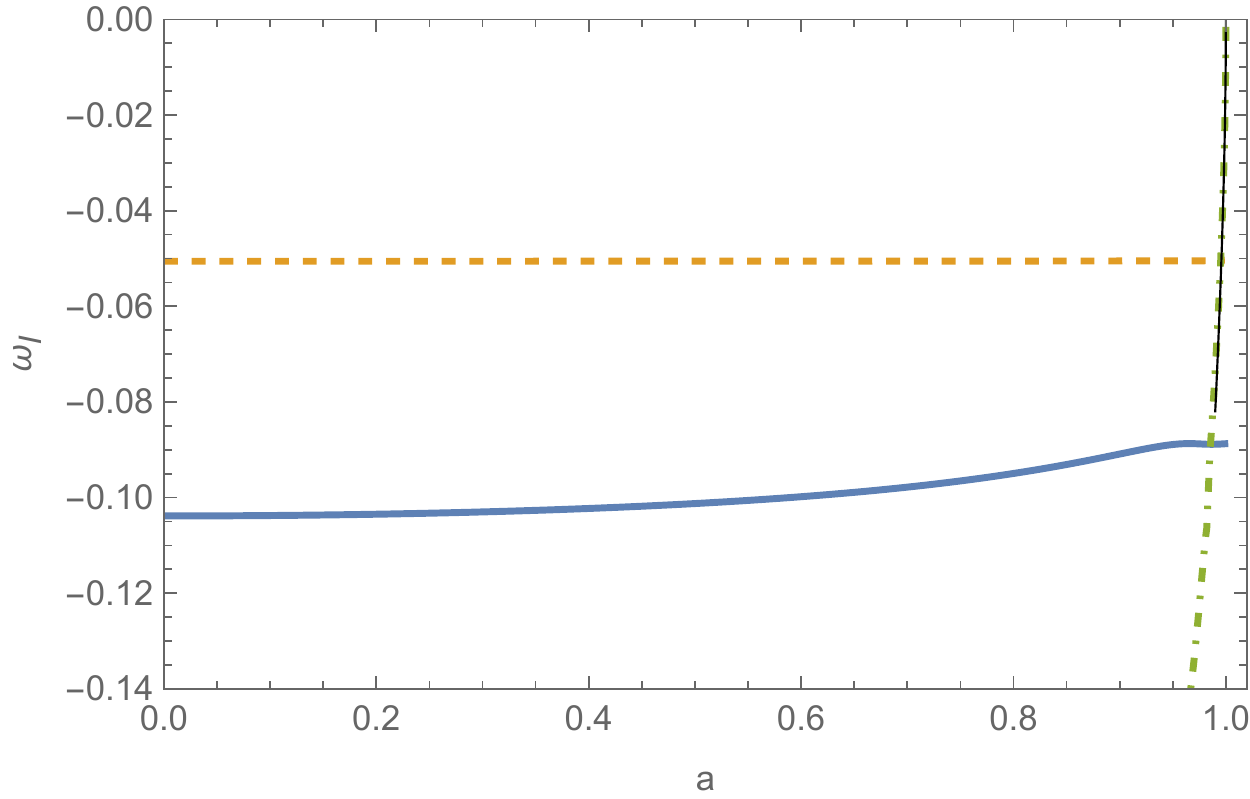}
    \includegraphics[width = 0.39\textwidth]{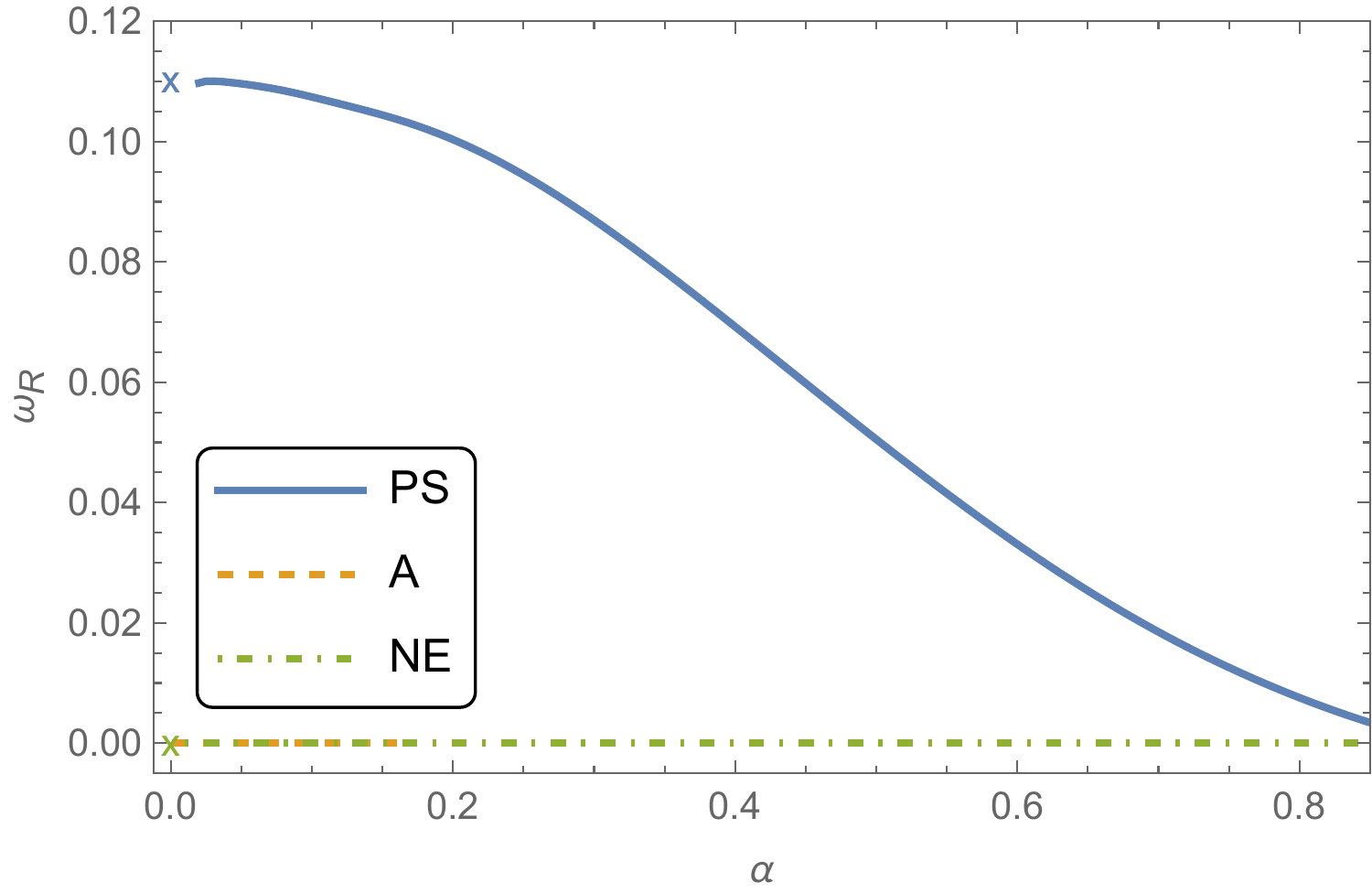}
    \includegraphics[width = 0.4\textwidth]{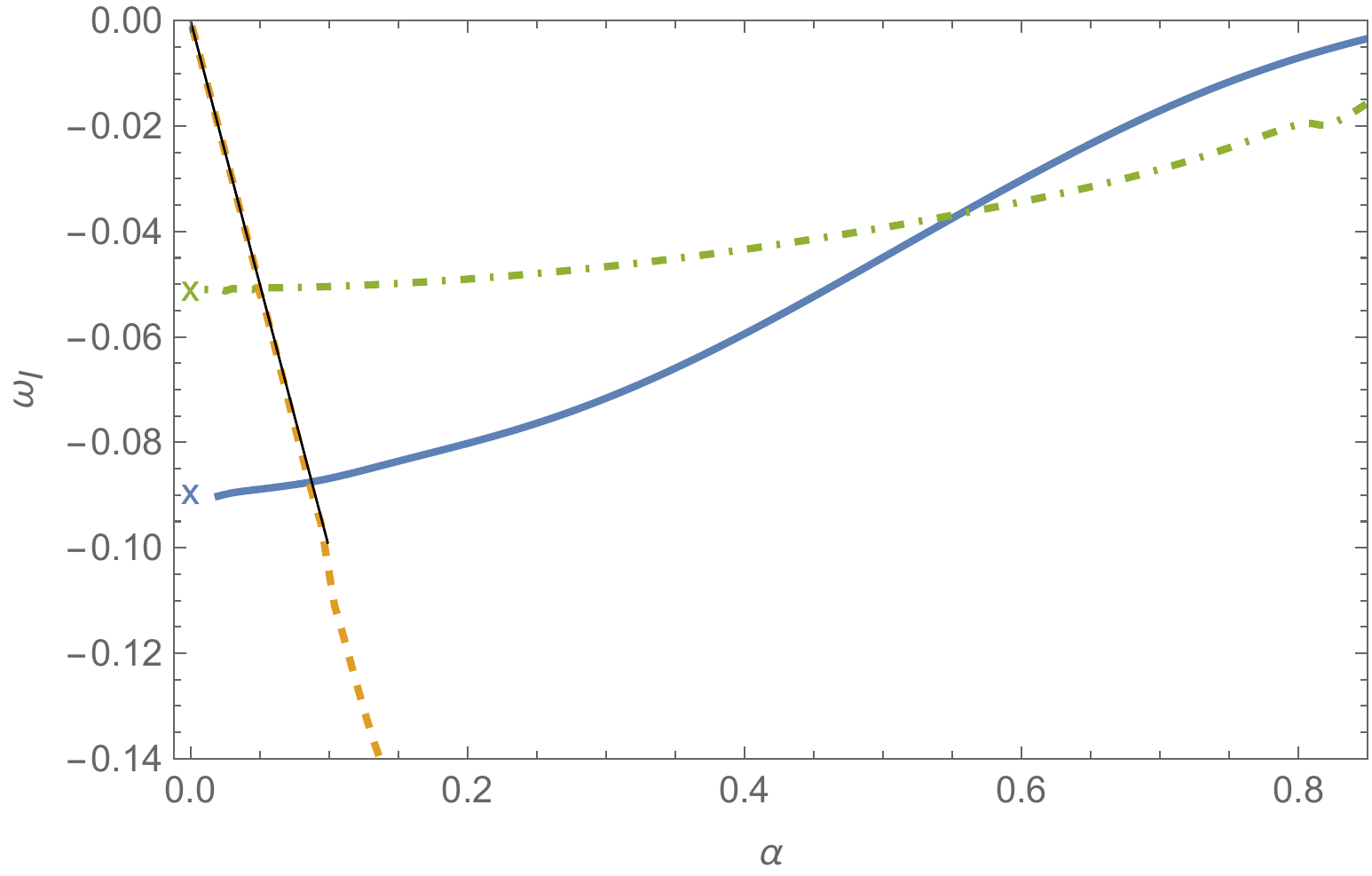}
    \caption{Real (left) and imaginary (right) parts of $n=0$ modes for both three families vs $a$ with  (upper) with $m_{0}=0,\alpha=0.05$ or vs $\alpha$ (bottom) with $m_{0}=0,a=0.995$. The markers "x" denote the QNM value in the Kerr limit ($\alpha=0$) determined by the Leaver's work. The black lines are calculated by the formula (\ref{eq:A modes}) in the small BH limit or (\ref{eq:NE modes1}) in the extreme BH limit, which provide the well approximant to QNMs.}
    \label{fig:three families}
\end{figure}

We have identified three families of scalar QNMs in the spinning C-metric, as depicted in Fig.\ref{fig:three families}, each with its unique physical origin. This QNM spectrum shares similarities with those observed in prior works of the charged C-metric\cite{Destounis:2020pjk} and the RN-dS BH\cite{Cardoso:2017soq}.

\textbf{Photon sphere (PS) modes $\mathbf{\omega_{\textrm{PS}}}$}: The photon sphere is defined as the trapping region of BHs where null particles are forced to travel in circular unstable orbits. In the eikonal limit ($m\sim l \gg 1$), the decay timescale $\omega_{\textrm{I}}$ of PS mode is directly related to the instability timescale of null geodesics near the photon sphere. The oscillation frequency $\omega_{\textrm{R}}$ of PS modes is associated with the orbital frequency of null geodesics \cite{Ferrari:1984zz,Cardoso:2008bp}. To maintain consistency in our notation, we refer to these modes as PS modes.

\begin{figure}[htbp]
    \centering
    \includegraphics[width = 0.45\textwidth]{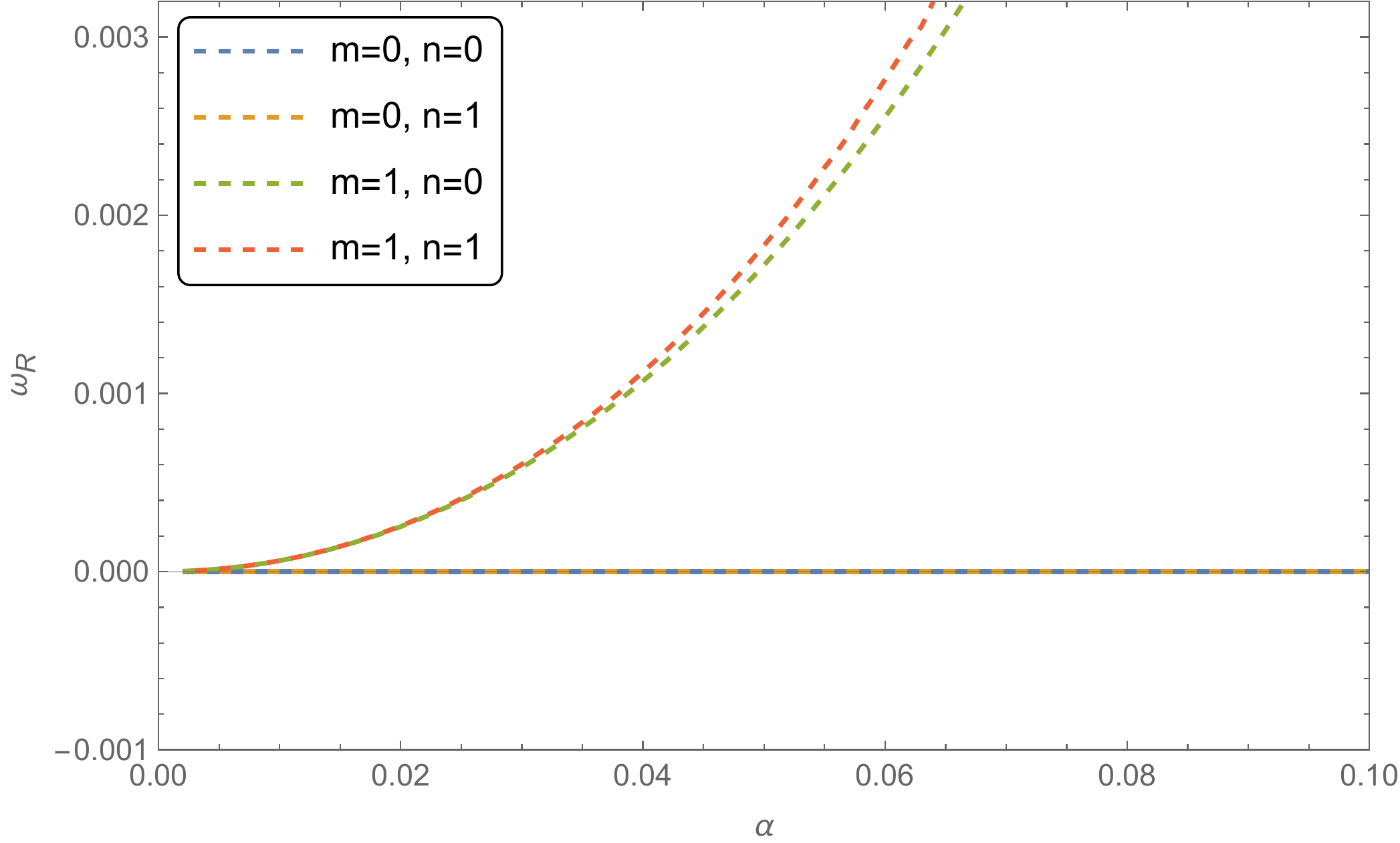}
    \includegraphics[width = 0.45\textwidth]{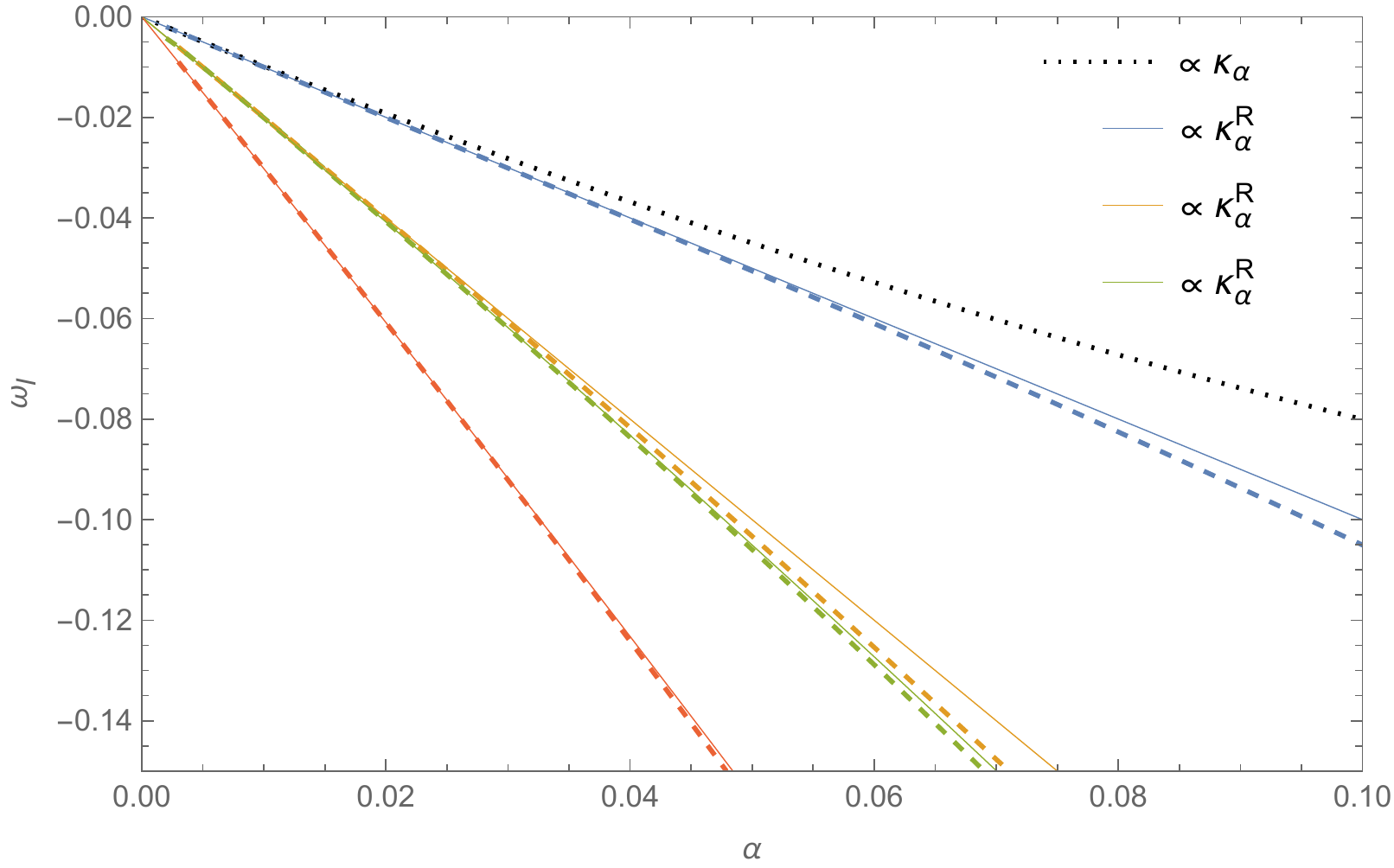}
    \caption{Real (left) and imaginary (left) part of acceleration modes  with different $n,m_{0}$ vs $\alpha$ on the background of rotating accelerating BH where $a=0.6$. The solid line is depicted by formula (\ref{eq:A modes}) and the dotted lines are using the same formula while replacing $\kappa_{\alpha}^{\textrm{R}}$ by $\kappa_{\alpha}$.}
    \label{fig:A vs alpha}
\end{figure}

\textbf{Acceleration (A) modes $\mathbf{\omega_{\textrm{A}}}$}: Acceleration modes exclusively arise from the damped perturbations in the boosted spacetime. The imaginary part of acceleration modes exhibits a linear relationship with the surface gravity at the acceleration horizon of the Rindler space
\begin{equation}
    \textrm{Im}(\omega_{A}) \simeq -\kappa^{\textrm{R}}_{\alpha} (m+n+1)i, \label{eq:A modes}
\end{equation}
with only a weak dependence on the rotation parameter. It is intriguing that the acceleration modes in a BH spacetime exhibit a dependence on $\kappa^{\textrm{R}}_{\alpha}$ rather than the surface gravity $\kappa_{\alpha}$ for the accelerating BH, which is first reported in \cite{Destounis:2020pjk}. The effect of the BH on these modes may be encoded in the modified azimuth number $m$ and its real part. We show this phenomenon in Fig.\ref{fig:A vs alpha}, where the solid line and the dotted line denote the linear dependence on $\kappa^{\textrm{R}}_{\alpha}$ and $\kappa_{\alpha}$ respectively. The intriguing dependence is also evident in the de Sitter modes for the RN-dS BH, which can be attributed to the role of the surface gravity at the cosmological horizon in governing the accelerated expansion of the RN-dS spacetime\cite{Cardoso:2017soq}. In Fig.\ref{fig:A vs alpha}, we have the acceleration modes with nonvanishing real part while the nonzero $m_{0}$ breaks the symmetry $\omega_{\textrm{R}} \rightarrow -\omega_{\textrm{R}}$.

\textbf{Near extreme (NE) modes $\mathbf{\omega_{\textrm{NE}}}$}: The third family of QNMs appears in the limit where the location of Cauchy horizon and event horizon tend to coincide. We also found a approximation while $a\rightarrow 1$
\begin{eqnarray}
    \textrm{Im}(\omega_{\textrm{NE}}) &\simeq& -\kappa_{-} (n+1)i,  \ \ \ \ \ \ \  m_{0} =0,1 \label{eq:NE modes1} \\
    \textrm{Im}(\omega_{\textrm{NE}}) &\simeq& -\kappa_{-} (n+1/2)i, \ \ \   m_{0} \geq 2. \label{eq:NE modes2}
\end{eqnarray}
Such modes, branching with the familiar damped modes beyond a critical value of $a$, have been found in the spectrum of Kerr BHs and were called zero damping modes. The approximation given in (\ref{eq:NE modes2}) is accurately consistent with the analytic prediction presented in \cite{Yang:2012pj,Yang:2013uba} with sufficient large $a$ and small $\alpha$. Remarkably, many similarities exist between the near extreme modes for the rotating Kerr BH and the charged version for the RN BH, although the physical interpretation of this correspondence remains unclear\cite{Cardoso:2017soq}. 

The fundamental mode (also called the dominant mode) corresponds to the slowest damped mode in the QNM spectrum. However, the presence of three distinct families of modes necessitates a distinguishable labeling convention. Specifically, for given parameters ($m_{0},a,\alpha$), we refer to the $n=0$ mode as the mode with the largest imaginary part for each of the three families, and label the overtones for each of the three families as $n=1,\ 2,\ \ldots$. We also adopt a convention of utilizing subscripts in the abbreviation for each QNM family (e.g., $\textrm{PS}_{0}$ to denote the $n=0$ photon sphere mode). The fundamental mode can be determined by comparing the $n=0$ modes of each QNM family.

Fig.\ref{fig:three families} illustrates the behavior of the $n=0$ modes for each family as a function of the parameter $a$ or $\alpha$. In the extreme BH limit ($a\rightarrow 1$), the near extreme modes become dominant in the QNM spectrum and tend to be on the real axis with a vanishing imaginary part, as demonstrated by the green line in the top right plot of Fig.\ref{fig:three families}. The black line is depicted from (\ref{eq:NE modes1}). On the other hand, in the limit $\alpha \rightarrow 0$, where the acceleration horizon tends to radial infinity, the acceleration modes become the fundamental modes. By comparing Fig.\ref{fig:three families} and Fig.\ref{fig:A vs alpha}, the acceleration modes are seen to emerge from the zero mode ($\omega = 0 $), which is related to a constant solution for radial equation (\ref{eq:radial}). The black line in the bottom right plot, calculated from (\ref{eq:A modes}), is well consistent with numerical results of acceleration modes. 

\begin{figure}[htbp]
    \centering
    \includegraphics[width = 0.45\textwidth]{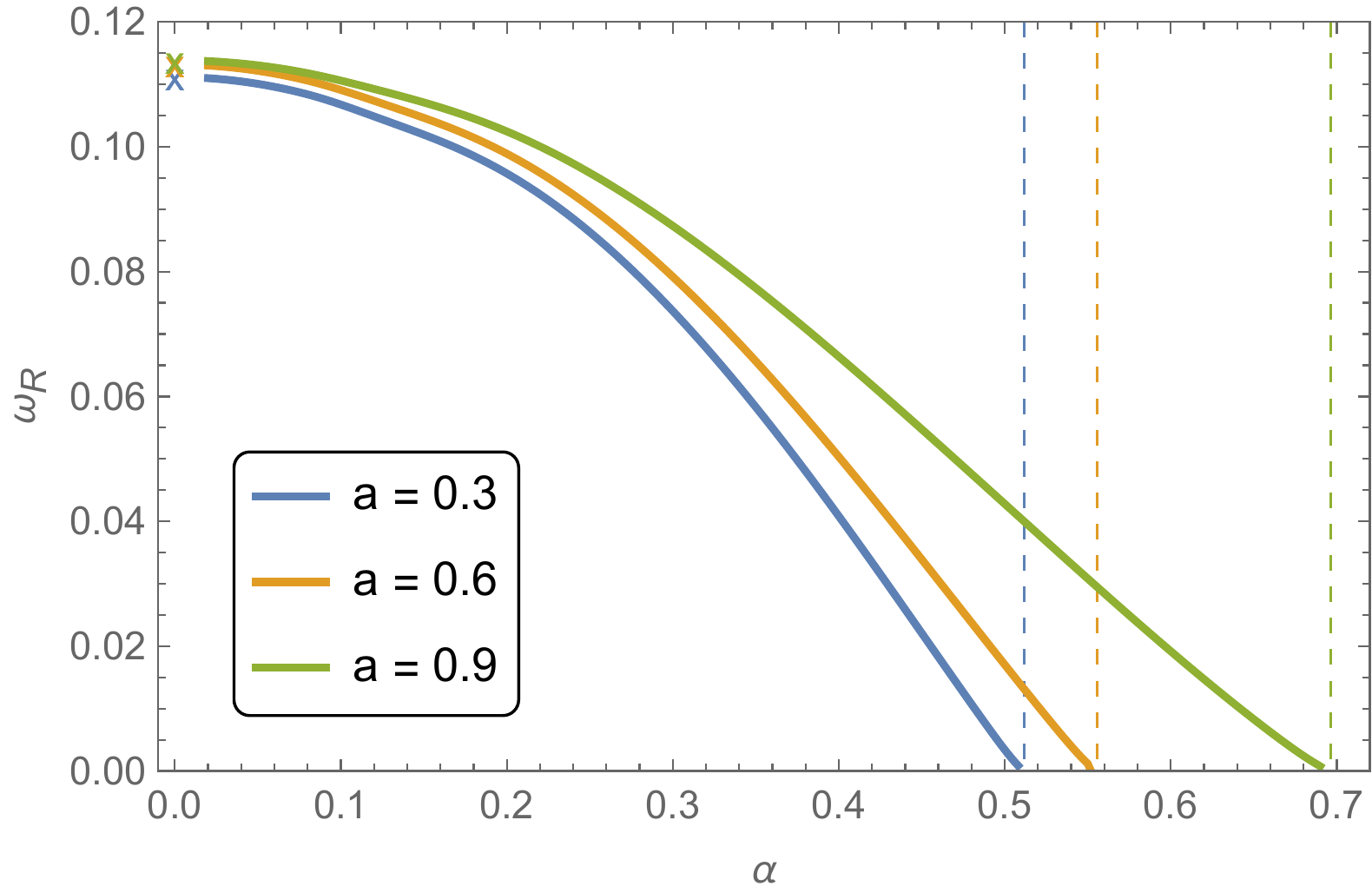}
    \includegraphics[width = 0.45\textwidth]{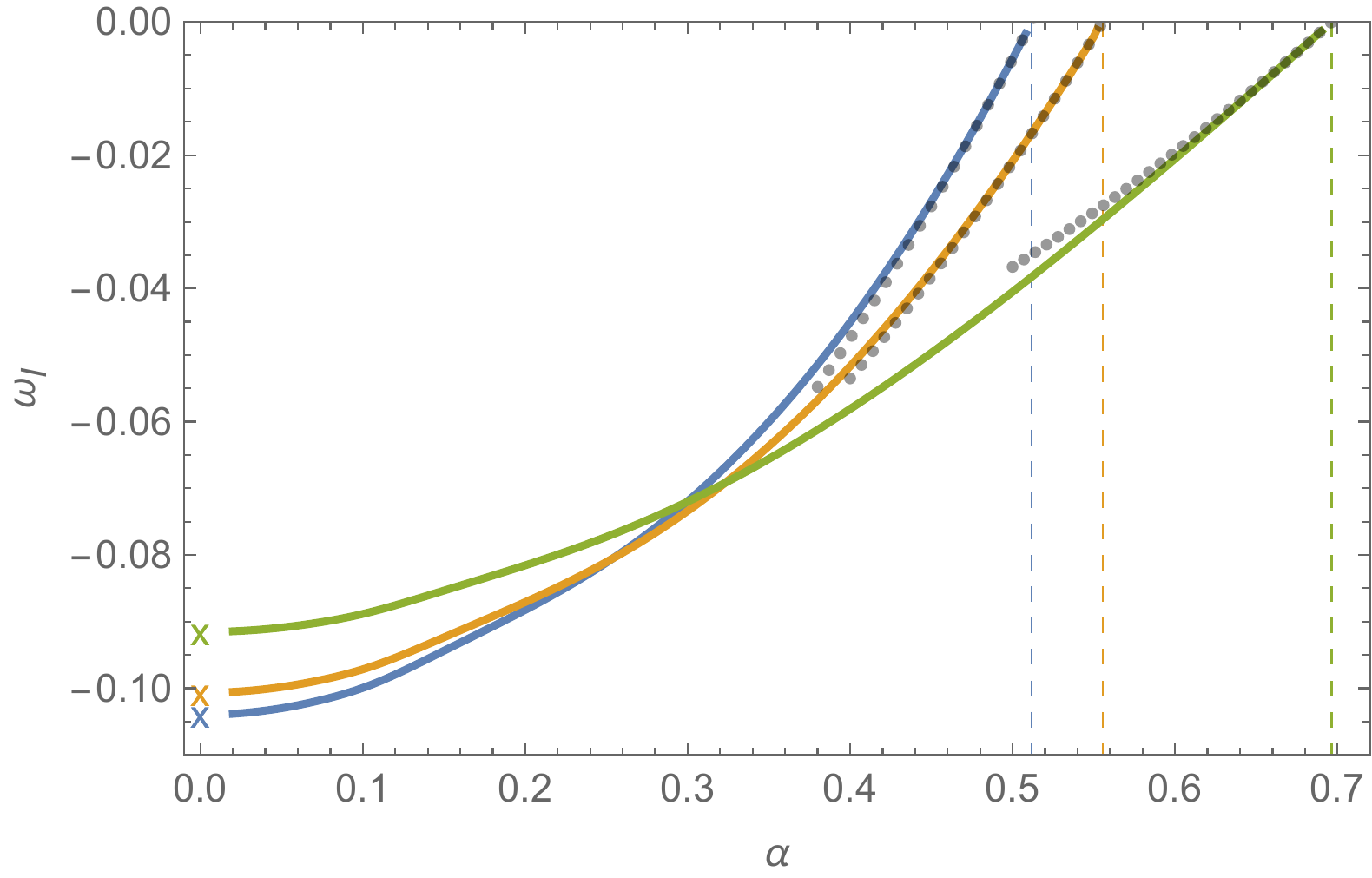}
    \caption{Real (left) and imaginary (right) parts of $\textrm{PS}_{0}$ modes vs $\alpha$ with $m_{0}=0$. The BH rotation parameter $a$ of different curves are shown in the frame. The dashed lines indicate the limit $\alpha_{\textrm{ext}}$. The markers "x" denote QNM values in the Kerr limit ($\alpha=0$) determined by the Leaver' work and The black dots are the analytic approximation $\textrm{Im}(\omega_{\textrm{PS}})=-(n+1/2)\kappa_{+}$ in the Nariai BH limit\cite{Gwak:2022nsi}.}
    \label{fig:PS vs alpha}
\end{figure}

\begin{figure}[htbp]
    \centering
    \includegraphics[width = 0.5\textwidth]{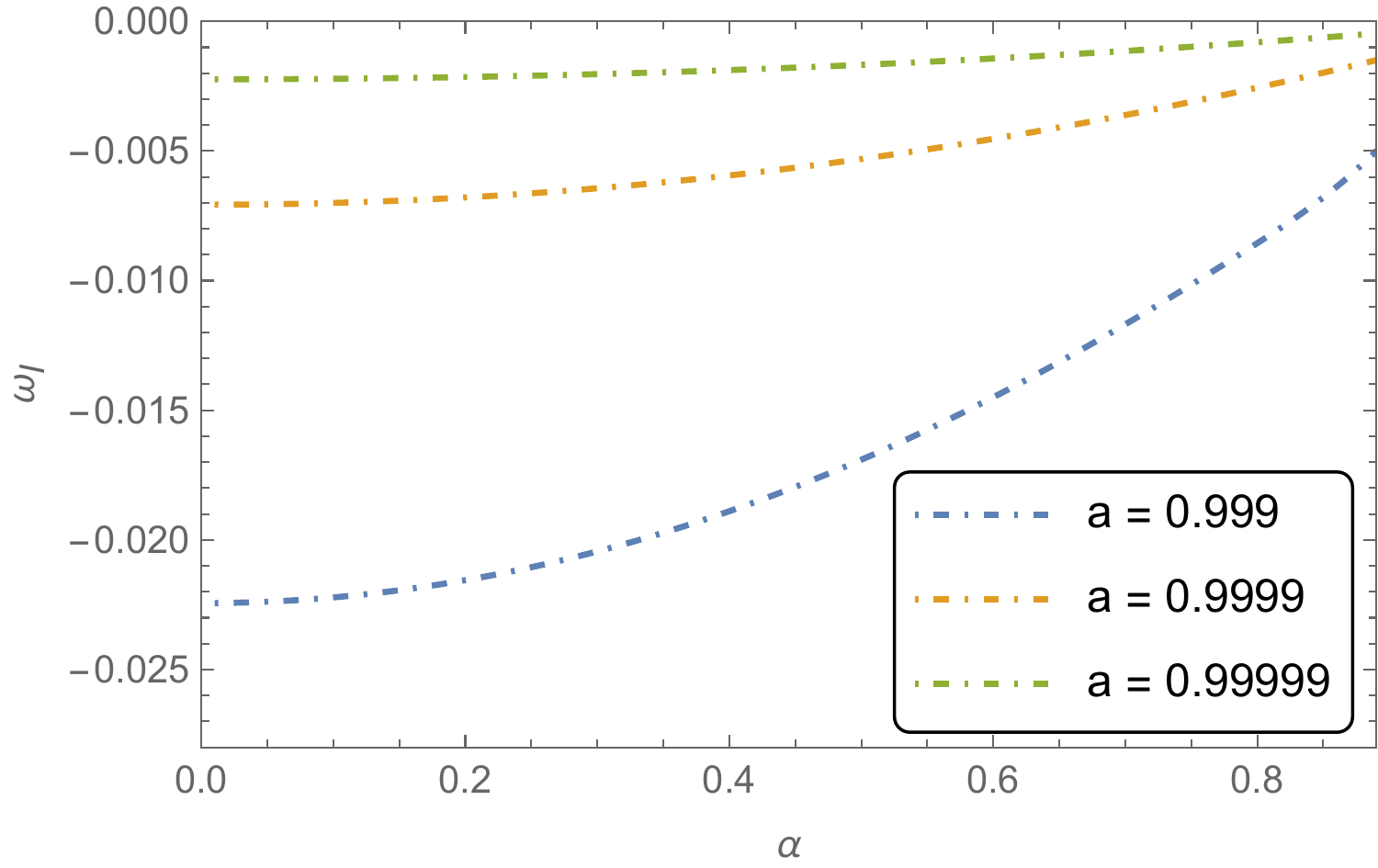}
    \caption{Imaginary parts of $n=0,m_{0}=0$ near extreme modes with different $a$ vs $\alpha$. }
    \label{fig:NE vs alpha}
\end{figure}

As $\alpha$ increases, the imaginary part of $\omega_{\textrm{PS}}$ increases more rapidly than that of the near extreme modes $\omega_{\textrm{NE}}$. In the limit $\alpha \rightarrow \alpha_{\textrm{ext}}$ where the event and acceleration horizon radius approach each other, as illustrated in Fig.\ref{fig:PS vs alpha}, PS modes become the zero mode characterized by vanishing real and imaginary parts. Similar trends are observed in Fig.\ref{fig:NE vs alpha}, where $\omega_{I}$ of the $\textrm{NE}_{0}$ mode increases with $\alpha$. However, due to the limitation of numerical calculation, we can not provide conclusive evidence as to whether or not the near extreme mode also becomes zero mode when $\alpha \rightarrow \alpha_{\textrm{ext}}$. 

\begin{figure}[htbp]
    \centering
    \includegraphics[width = 0.5\textwidth]{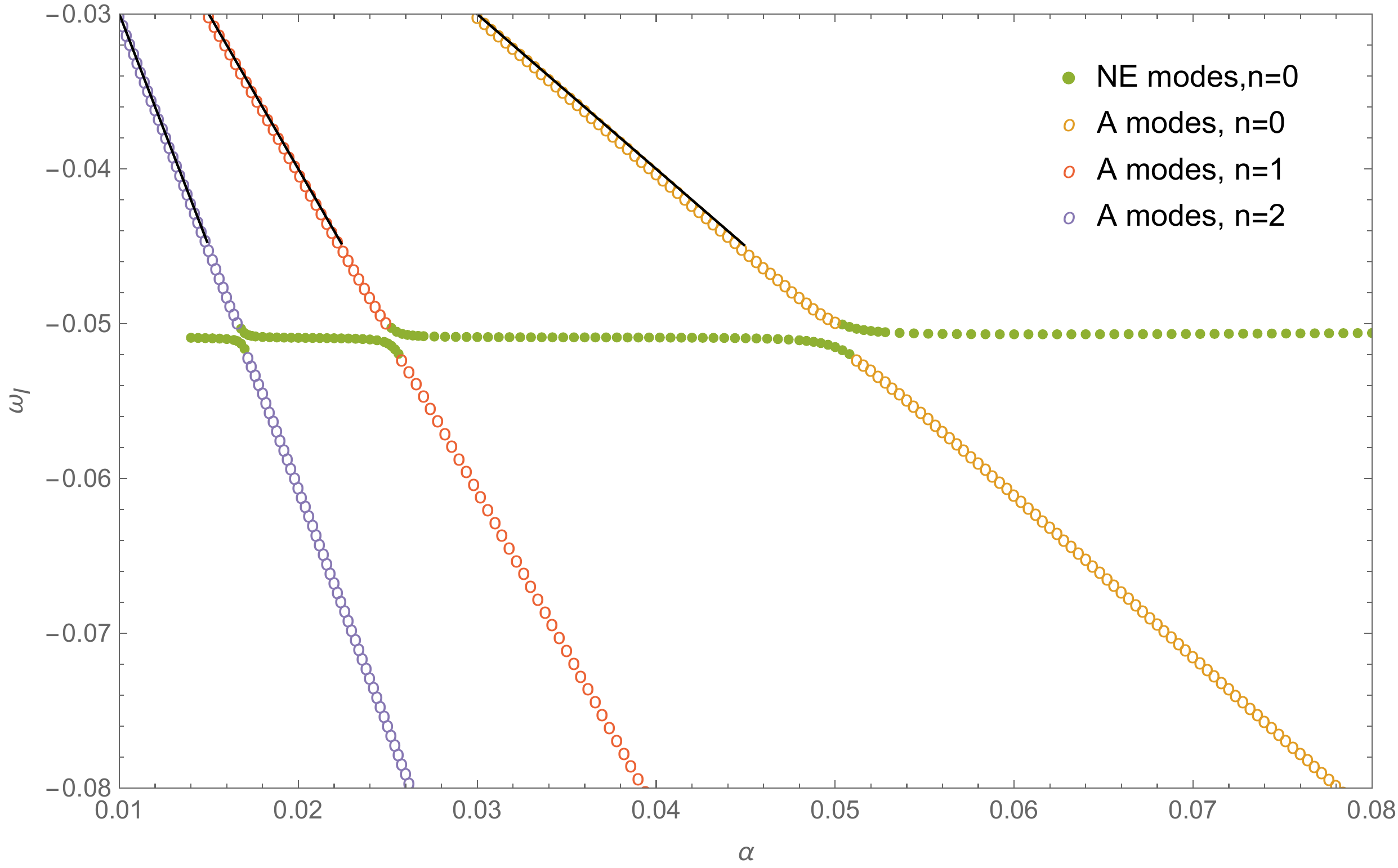}
    \caption{The eigenvalue repulsion of $\textrm{A}_{i}$ families and $\textrm{NE}_{0}$ modes with $m_{0}=0,\ a=0.995$. There are frequency gaps in the neighbourhood of two kinks of the two corresponding curves generating from the connection between $\textrm{A}_{i}$ modes (for any $i=0,1,\ldots$) and $\textrm{NE}_{0}$ modes. The black lines are depicted by formula (\ref{eq:A modes}).}
    \label{fig:eigenvalue repulsion 1}
\end{figure}

A new phenomenon, called eigenvalue repulsion, appears in the bottom right plot of Fig.\ref{fig:three families} while $\textrm{A}_{0}$ line intersects with $\textrm{NE}_{0}$ line. Further examination of this phenomenon is provided in Fig.\ref{fig:eigenvalue repulsion 1}. The acceleration modes exhibit a rapid decrease with increasing $\alpha$, while the $\textrm{NE}_{0}$ mode varies slightly. As a crossover of two lines occurs, both the $\textrm{A}_{0}$ line (yellow) and the $\textrm{NE}_{0}$ line break into two pieces and connect with each other. Consequently, a gap emerges between the two kinks of the two new curves, diminishing the distinction between the different families. However, the black line calculated using (\ref{eq:A modes}) can still serve to distinguish between the two modes. Similar phenomenons are observed while $\textrm{A}_{1}$ line and $\textrm{A}_{2}$ line intersect with the $\textrm{NE}_{0}$ line. 

\begin{figure}[htbp]
    \centering
    \includegraphics[width = 0.45\textwidth]{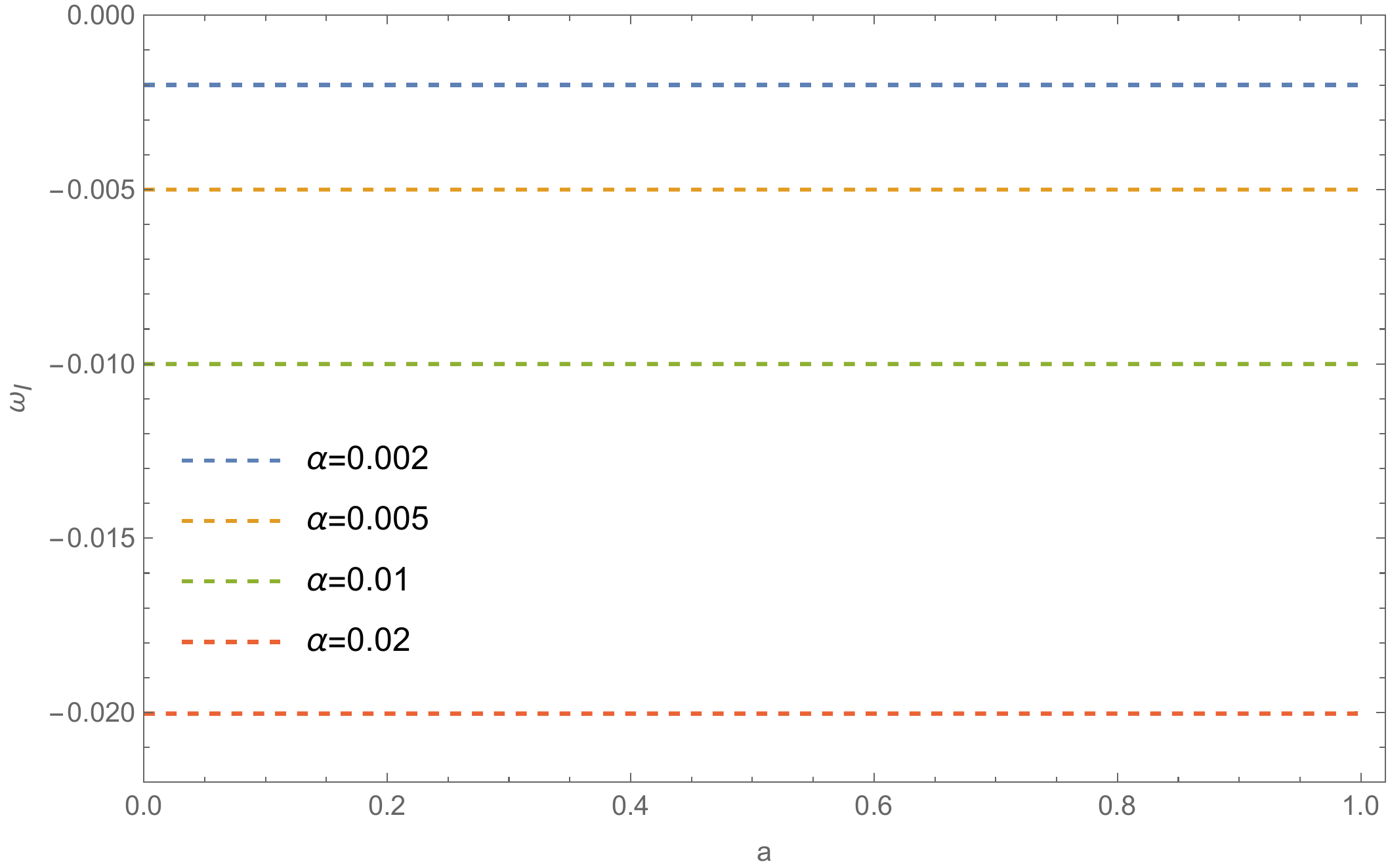}
    \includegraphics[width = 0.45\textwidth]{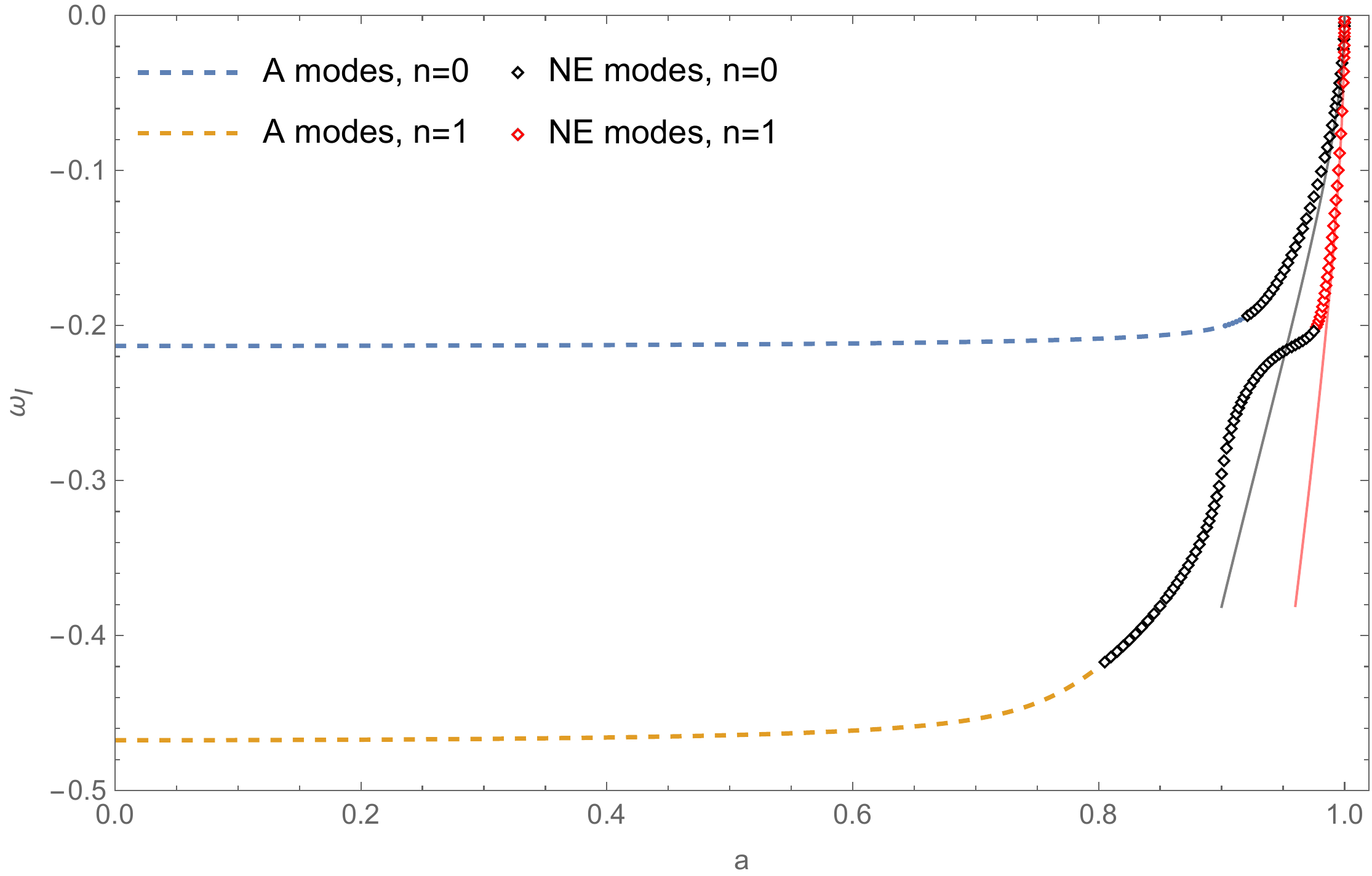}
    \caption{The $n=0,m_{0}=0$ acceleration modes with different $\alpha$ vs $a$. The right plot depicts the eigenvalue repulsion between $\textrm{A}_{0}$ modes, $\textrm{A}_{1}$ modes, $\textrm{NE}_{0}$ modes and $\textrm{NE}_{1}$ modes while $m_{0}=0,\  \alpha=0.2$. The solid lines are evaluated by (\ref{eq:NE modes1}) respectively. The eigenvalue repulsion can also exist in the left plot while $a$ is sufficiently large to bring the line of NE families cross these $\textrm{A}_{0}$ line. }
    \label{fig:A vs a}
\end{figure}

In the right plot of Fig.\ref{fig:A vs a}, a more intricate type of eigenvalue repulsion is presented. This phenomenon arises when the acceleration modes encounter the increasing near extreme modes at the imaginary axis while $a$ increases. The $\textrm{NE}_{0}$ line breaks into two pieces and the upper branch of this line connects smoothly to the $\textrm{A}_{0}$ family. Below, a continuous curve exists in which the bottom branch of $\textrm{NE}_{0}$ line bridges the $\textrm{A}_{1}$ line and the upper branch of $\textrm{NE}_{1}$ line. Similar curves connecting the $\textrm{A}_{i}$ line, $\textrm{NE}_{i-1}$ line and $\textrm{NE}_{i}$ line in a sequential manner ($i=1,\ 2,\ \ldots$) also exist although they are not depicted in Fig.\ref{fig:A vs a}. The distinctions between different families or overtones are destroyed by the eigenvalue repulsion. Before the eigenvalue repulsion occurs, we found that the acceleration modes are insensitive to $a$.

\subsection{Strong cosmic censorship conjecture}

\begin{figure}[htbp]
    \centering
    \includegraphics[width = 0.5\textwidth]{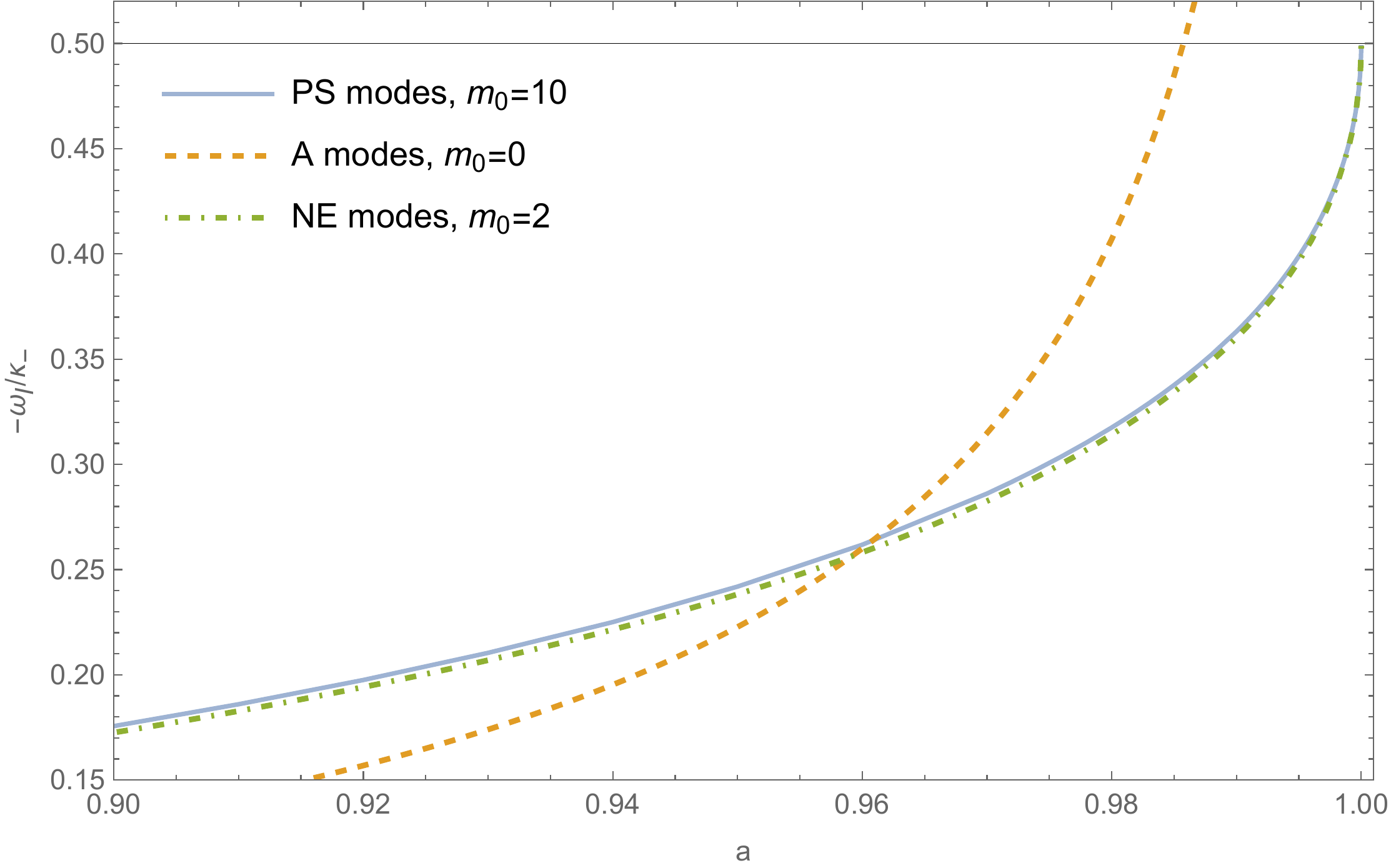}
    \caption{$-\textrm{Im}(\omega)/\kappa_{-}$ for each families of QNMs with $\alpha=0.05$. The $\beta$ is obtained by extracting the minimum value among three families.}
    \label{fig:beta}
\end{figure}

We follow the derivation from the case of Kerr-dS BHs presented in \cite{Dias:2018ynt}. One can convert the coordinates $(t,r,\theta,\phi)$ into the outgoing coordinates $(u,r,\theta,\phi')$ where the definition of new coordinates is given by
\begin{equation}
    dt=du+\frac{r^{2}+a^{2}}{Q} dr , \ \ \ d\phi = d\phi'+\frac{a}{Q} dr.
    \label{eq:outgoing coordinate}
\end{equation}
Then the corresponding separable solution is imposed by 
\begin{equation}
    \varphi(u,r,\theta,\phi') = \Omega \  e^{-i\omega u} e^{i m \phi'} R(r) \frac{Y(\theta)}{\sqrt{P}}.
\end{equation}
Now there are two independent solutions near the Cauchy horizon
\begin{eqnarray}
    \varphi^{(1)} &=& \Omega \ e^{-i\omega u} e^{i m \phi'} R^{(1)}(r) Y(\theta)/\sqrt{P},   \\
    \varphi^{(2)} &=& \Omega \ e^{-i\omega u} e^{i m \phi'} (r-r_{-})^{i[\omega-am/(r_{-}^{2}+a^{2})]/\kappa_{-}} R^{(2)}(r) Y(\theta)/\sqrt{P},  
    \label{eq:independent solutions}
\end{eqnarray}
with some non-zero functions $R^{(1)}(r)$ and $R^{(2)}(r)$. The regularity of QNMs is determined by the non-smooth solution $\varphi^{(2)}$. The violation of SCC requires $\varphi^{(2)}$ to be locally square integrable, i.e., the condition that $(\partial_{r}\varphi^{(2)})^{2}$ is integrable gives 
\begin{equation}
    \beta>1/2 \ \ \ \ \  \textrm{where} \ \ \ \ \ \beta \equiv -\frac{\textrm{Im}(\bar{\omega})}{\kappa_{-}},
    \label{eq:beta}
\end{equation}
which $\bar{\omega}$ denotes the fundamental modes in the spectrum. In other words, if a QNM $\omega$ with $-\textrm{Im}(\omega)/\kappa_{-}<1/2$ is discovered, one can determine that the SCC conjecture is respected under such perturbation for the given parameters ($\alpha$, $a$). We depict $-\textrm{Im}(\omega)/\kappa_{-}$ for each family of QNMs in Fig.\ref{fig:beta} and demonstrate that the minimum value among three families is always below $1/2$. These results can be expected from (\ref{eq:NE modes2}), where the imaginary part of near extreme modes with $n=0$, $m_{0}\geq 2$ are proportional to $-i\kappa_{-}/2$ while $a \rightarrow 1$. On the other hand, the charged version of near extreme modes with $n=l=0$ is proportional to $-i\kappa_{-}$ in RN-dS BHs or charged C-metric\cite{Cardoso:2017soq,Destounis:2020pjk}. The different behaviors of modes in the extreme BH limit lead to different fates of SCC between charged and rotating BHs. 

\subsection{Comparison of methods}

\label{secIV.3}

\begin{table}[h]
    \centering
    \scalebox{0.8}{
    \begin{tabular}{|c|c|c|c|} 
        \hline  
        \ \ parameters \ \  &\ families \  & \ \ \ continued fractions method \ \ \ & \ \ \  directly integration method \ \ \ \\ [0.5ex] 
        \hline 
        $\alpha=0.03$ & $\textrm{PS}_{0}$ & $0.30327363 - 0.09736685 i$ & $0.30327413 - 0.09736639 i$ \\ 
        $a=0$ & $\textrm{A}_{0}$ & $-0.06189986 i$ & $-0.06189984 i$ \\
        $m_{0}=1$ & $\textrm{NE}_{0}$ & $--$ & $--$ \\
        \hline 
        $\alpha=0.2$ & $\textrm{PS}_{0}$ & $0.09888222 - 0.08707878 i$ & $0.09888222 - 0.08707878 i$ \\ 
        $a=0.6$ & $\textrm{A}_{0}$ & $-0.21156725 i$ & $-0.21156725 i$ \\
        $m_{0}=0$ & $\textrm{NE}_{0}$ & $--$ & $--$ \\
        \hline
        $\alpha=0.1$ & $\textrm{PS}_{0}$ & $0.10742075 - 0.08685416 i$ & $0.10742077 - 0.08685417 i$ \\ 
        $a=0.995$ & $\textrm{A}_{0}$ & $-0.10804081 i$ & $-0.10804147 i$ \\
        $m_{0}=0$ & $\textrm{NE}_{0}$ & $-0.05043962 i$ & $-0.05043962 i$ \\
        \hline
        $\alpha=0.05$ & $\textrm{PS}_{0}$ & $--$ & $--$ \\ 
        $a=0.99998$ & $\textrm{A}_{0}$ & $-0.05055091 i$ & $-0.05033569 i$ \\
        $m_{0}=0$ & $\textrm{NE}_{0}$ & $ -0.00315480 i$ & $-0.00315558 i$ \\
        \hline
        $\alpha=0.9$ & $\textrm{PS}_{0}$ & $0.00187795 - 0.00176789 i$ & $0.00187795 - 0.00176789 i$ \\ 
        $a=0.999$ & $\textrm{A}_{0}$ & $--$ & $--$ \\
        $m_{0}=0$ & $\textrm{NE}_{0}$ & $-0.00453841 i$ & $-0.00453841 i$ \\
        \hline
        $\alpha=0.15$ & $\textrm{PS}_{0}$ & $0.10113821 - 0.09480198 i$ & $0.10113821 - 0.09480198 i$ \\ 
        $a=0.1$ & $\textrm{A}_{0}$ & $-0.16337640 i$ & $ -0.16337637 i$ \\
        $m_{0}=0$ & $\textrm{NE}_{0}$ & $--$ & $--$ \\
        \hline
    \end{tabular}}
    \caption{The comparison between two methods with different parameters. We displays only the $n=0$ mode for each families. There are some modes not provided because they are higher overtones at such parameter values, which is difficult to be determined by numerical methods. As shown in this table, the two method generally provide more consistent and accurate results for lower overtones. The first row with the separation constant $\lambda=-1.09236715$ regains the results given in \cite{Destounis:2020pjk},  where the separation constant requires a transformation $\lambda' = -2 \lambda+1/3$. }
    \label{table1}
    \end{table}

    \begin{table}[h]
        \centering
        \scalebox{0.8}{
        \begin{tabular}{|c|c|c|c|c|} 
            \hline  
           \ \ \ \ parameters \ \ \ \  & \ families \  & \ \ \ continued fractions method \ \ \ & \ \ \  directly integration method \ \ \ & \ \  approximation \ \ \\ [0.5ex] 
            \hline 
            $\alpha=0.1$ & $\textrm{NE}_{0}$ & $1.20792690 - 0.00034947 i$ & $--$ & $-0.00035052 i$  \\
            $a=0.999999$ & $\textrm{NE}_{1}$ & $1.20792630 - 0.00104781 i$ & $--$ & $-0.00105157 i$ \\
            $m_{0}=2$ & $\textrm{NE}_{2}$ & $1.20792713 - 0.00174452 i$ & $--$ & $-0.00175262 i$ \\
            \hline
            $\alpha=0.05$ & $\textrm{NE}_{0}$ & $-0.00022301 i$ & $-0.00022617 i$ & $-0.00022315 i$  \\
            $a=0.9999999$ & $\textrm{NE}_{1}$ & $-0.00044604 i$ & $-0.00036142 i$ & $-0.0004463 i$ \\
            $m_{0}=0$ & $\textrm{NE}_{2}$ & $-0.00066907 i$ & $--$ & $-0.00066944 i$ \\
            \hline
            $\alpha=0.002$ & $\textrm{A}_{0}$ & $--$ & $-0.00200002 i$ & $-0.002 i$  \\
            $a=0.5$ & $\textrm{A}_{1}$ & $--$ & $-0.00399363 i$ & $-0.004 i$ \\
            $m_{0}=0$ & $\textrm{A}_{2}$ & $--$ & $-0.00599757 i$ & $-0.006 i$ \\
            \hline
            $\alpha=0.508$ & $\textrm{PS}_{0}$ & $--$ & $0.00088583 - 0.0018015  i$ & $-0.00179944 i$  \\
            $a=0.3$ & $\textrm{PS}_{1}$ & $--$ & $0.00092112 - 0.0053762 i$ & $-0.00539831 i$ \\
            $m_{0}=0$ & $\textrm{PS}_{2}$ & $--$ & $0.00054932 - 0.00908775 i$ & $-0.00899718 i$ \\
            \hline
        \end{tabular}}
        \caption{The comparison between QNMs and the approximants from (\ref{eq:A modes}), (\ref{eq:NE modes1}), (\ref{eq:NE modes2}) or the analytic formula $\textrm{Im}(\omega_{\textrm{PS}})=-(n+1/2)\kappa_{+}$ presented in \cite{Gwak:2022nsi}. The approximants are only shown its imaginary parts. Some modes were not provided due to the limitations of the corresponding numerical methods.}
        \label{table2}
        \end{table}

    The table.\ref{table1} shows QNMs calculated by the continued fractions method and the directly integration method respectively. The two methods provide precisely consistent results. We also present QNMs with some extreme parameter choices (e.g., $a\rightarrow 1$ for the first two rows, $\alpha\rightarrow 0$ for the third row, and $\alpha \rightarrow \alpha_{\textrm{max}}$ for the fourth row) in table.\ref{table2}. At such limit, only one method can give effective results, and its imaginary part matches the corresponding approximant. Another method becomes numerically unstable or provides unreliable discontinuous results. We show that the computable parameter spaces of the two methods are complementary.

\section{Discussion}

\label{sectionV}

This study focuses on the scalar ($s=0$) QNM spectrum of rotating accelerating BHs calculated numerically by two methods (the continued fractions method and the directly integration method). Three families of QNMs are identified, namely photon sphere modes, acceleration modes, and near extreme modes. We examine the dependence of each family on various parameters, such as BH rotation and acceleration. We found that the acceleration modes demonstrate a linear dependence with the surface gravity $\kappa_{\alpha}^{\textrm{R}}$ at the acceleration horizon of Rindler space in the small BH limit ($\alpha \rightarrow 0$), while the near extreme modes are consistent with (\ref{eq:NE modes1}) in the extreme BH limit ($a\rightarrow 1$). The photon sphere modes become dominant and tend to zero mode while $\alpha$ approach its extreme value $\alpha_{\textrm{ext}}$. Our results are reliable because of the good agreements of the comparison between the two methods or the analytic approximations in previous works. We also discuss the eigenvalue repulsion phenomenon that occurs when the acceleration modes intersect with the near extreme modes at certain values of parameters. The distinction between different families of modes is diminished or even destroyed by the eigenvalue repulsion. The SCC conjecture is determined by $\beta$. We found no evidence of SCC violation. 

The gravitational ($s=-2$) and electromagnetic ($s=-1$) perturbation of rotating accelerating BH are still the opening problems. The precise dependence of acceleration modes on the surface gravity in the Rindler space suggests that these modes may exist beyond the spinning C-metric with axial symmetry. We deduce that, for rotating BHs or even compact objects without event horizon with arbitrary acceleration directions, the imaginary part of their acceleration modes should be approximately equal to the numerical results presented in this paper. When the BH has a sufficient small acceleration, such modes become the dominant or even the fundamental modes in the gravitational wave spectrum.  Information about BHs, such as their spin and acceleration direction, may be encoded in the real part of their acceleration modes. Therefore, providing a ringdown template may enable us to search for moving and accelerating BHs in gravitational wave signals. This then offers another way alternative to the gravitational lensing \cite{Ashoorioon:2022zgu} to distinguish slowly accelerating BHs.

\begin{acknowledgments}
	The work is in part supported by NSFC Grant
	No.12205104 and the startup funding of South China University of
	Technology.
\end{acknowledgments}

\end{document}